# High-resolution studies of the Majorana atomic chain platform

Benjamin E. Feldman[*], Mallika T. Randeria[*], Jian Li[*], Sangjun Jeon, Yonglong Xie, Zhijun Wang, Ilya K. Drozdov[*†], B. Andrei Bernevig, Ali Yazdani[‡]

*Joseph Henry Laboratories and Department of Physics, Princeton University, Princeton, New Jersey 08544, USA*

[†]Present address: *Condensed Matter Physics and Materials Science Department, Brookhaven National Laboratory, Upton, New York 11973, USA*

[*]These authors contributed equally to the manuscript.
[‡]email: yazdani@princeton.edu

**Abstract**: Ordered assemblies of magnetic atoms on the surface of conventional superconductors can be used to engineer topological superconducting phases and realize Majorana fermion quasiparticles (MQPs) in a condensed matter setting. Recent experiments have shown that chains of Fe atoms on Pb generically have the required electronic characteristics to form a 1D topological superconductor and have revealed spatially resolved signatures of localized MQPs at the ends of such chains. Here we report higher resolution measurements of the same atomic chain system performed using a dilution refrigerator scanning tunneling microscope (STM). With significantly better energy resolution than previous studies, we show that the zero bias peak (ZBP) in Fe chains has no detectable splitting from hybridization with other states. The measurements also reveal that the ZBP exhibits a distinctive 'double eye' spatial pattern on nanometer length scales. Theoretically we show that this is a general consequence of STM measurements of MQPs with substantial spectral weight in the superconducting substrate, a conclusion further supported by measurements of Pb overlayers deposited on top of the Fe chains. Finally, we report experiments performed with superconducting tips in search of the particle-hole symmetric MQP signature expected in such measurements.

**Introduction**

Condensed matter systems provide a versatile platform for the realization of emergent phases that host exotic quasiparticles and exhibit novel electronic phenomena. Recently, there has been considerable interest in material systems in which superconductivity has a topological nature and Majorana fermion quasiparticle (MQP) excitations emerge either as edge modes or as core states of field-induced vortices[1-3]. The search for a solid state realization of MQPs has been motivated both by the possible discovery of a neutral fermion that is its own antiparticle, as first envisioned by Ettore Majorana[4], and by the prediction that solid state MQPs will obey non-Abelian statistics. The latter can be used to realize topological qubits for fault-tolerant quantum computation[5, 6]. Initial searches for MQPs involved strongly interacting electron systems, such as fractional quantum Hall states[7-10]. However, more recent efforts have focused on engineering topological superconductivity by combining conventional BCS superconductors and spin textured electronic systems, such as the surface states of a topological insulator[11], Rashba semiconducting nanowires[12, 13], or assemblies of magnetic atoms[14-24]. The key idea is that the spin texture of the electronic bands in these material platforms stabilizes an effective p-wave pairing through the proximity effect, and produces localized MQPs at the boundaries of the structure.

To date, strong evidence for the presence of MQPs has come from transport studies of hybrid superconductor-semiconductor nanowire devices[25], from scanning tunneling microscopy (STM) measurements of chains of magnetic atoms on the surface of a superconductor[23], and from STM measurements of vortices in superconductor-topological insulator heterostructures[26]. In proximitized semiconducting nanowires with strong spin-orbit interaction, transport studies showed a zero bias peak (ZBP) for a range of parallel magnetic fields and gate voltages[25, 27-29] as well as the fractional ac Josephson effect[30]. This is consistent with the appearance of a MQP when the field drives these nanowires into the topological phase. More recent experiments demonstrated the expected change in charging that is characteristic of such a nanowire in the topological phase, as well as experimental signatures of exponentially suppressed coupling between MQP end modes[31]. A recent effort aims to discount alternative explanations for the ZBP, such as the Kondo effect or disorder[32-36], by examining cleaner nanowire devices[37].

Chains of magnetic atoms on the surface of a superconductor provide a novel approach to engineer topological superconductivity with the unique advantage that MQPs can be directly visualized using spatially resolved STM measurements. Previous spin-polarized STM studies of self-assembled chains of Fe atoms on superconducting Pb showed clear evidence for both ferromagnetic ordering of the Fe atoms and strong spin-orbit coupling at the surface of the Pb substrate[23]. The combination of these two ingredients together with proximity-induced superconductivity is predicted to almost always produce a topological superconducting phase in such atomic chains[24]. Consistent with this prediction, spatially resolved spectroscopic measurements showed the presence of a ZBP in the local density of states (LDOS) at the end of the chains, as expected for a localized MQP[23]. Subsequent experiments have probed the properties of Fe chains on Pb with superconducting tips and have reported similar zero-energy states as well as explored other features on this platform[38, 39].

The ZBP in these chains was shown not to be due to the Kondo effect as it is fully suppressed when a small magnetic field (well below the Kondo temperature) is applied to drive the Pb substrate normal. However, other questions regarding the MQP interpretation of the ZBP for atomic chains require more detailed investigation. Initially, the nanometer length scale reported for the localization of MQPs from spectroscopic mapping was suggested to be inconsistent with the small predicted p-wave gap[40]. Subsequent theoretical efforts from several groups, however, have shown that the short MQP localization length is expected due to strong velocity renormalization in this hybrid system; the characteristic MQP localization length is much shorter than the coherence length of the host superconductor[24, 41]. Another concern is the temperature at which the previous measurements were performed (around 1 K), which is comparable to the expected size of the p-wave gap. The temperature limited the precision with which previous experiments could determine the possible splitting of the ZBP. A very accurate energy resolution could distinguish a MQP from a pair of states very close to zero energy, and could also enable detailed examination of the background of low energy electronic states.

The focus of the current study is to extend the previous experiments to a lower temperature that allows us to probe the magnetic atomic chains with much higher energy resolution. We demonstrate that the ZBP in atomic Fe chains remains pinned at zero energy with no detectable hybridization with other electronic states, and we reveal new information about the background of low-energy quasiparticle (Shiba[42-44]) states. We also provide the first experimental evidence that the ZBP has substantial weight in the host superconductor based on the observation of a characteristic 'double eye' spatial pattern of the zero-bias conductance, as well as measurements of Fe chains that have been covered with a monolayer of Pb. We develop theoretical models that accurately describe both the detailed nature of the states within the Fe chains as well as the observation of enhanced MQP weight in the superconducting substrate. Finally, we describe experiments using superconducting tips at lower temperatures motivated by a recent proposal for detecting the particle-hole symmetry of MQPs[39, 45]. Taken together, these results provide significant evidence in support of the MQP interpretation of the ZBP in the Fe/Pb atomic chain system.

**High Resolution Spectroscopy and Spatial Structure of the Zero Mode**

We extend the previous experiments to lower temperature using a dilution refrigerator STM[46], which cools samples to about 20 mK. We fabricate atomically ordered one-dimensional Fe chains on a pristine Pb(110) surface as described in ref. 23, yielding chains up to 40 nm in length with pristine portions ranging from 5-20 nm (Fig. 1a). Measurements of the LDOS with normal tips far from the chains reveal spectral features associated with the two energy gaps expected from the two Fermi surfaces of Pb[47], and the data fit well to theory using an electron temperature of 250 mK (Fig. 1b). The majority of the Fe chains that we have explored show a prominent ZBP in spectroscopic measurements near their end (Fig. 1c) and display no sign that it is offset from zero bias or split by hybridization with other states, with an experimental upper bound of 80 µeV splitting. As shown in Fig. 1d, the full width at half maximum of the ZBP can be as low as 90 µeV (above the background measured in the middle of the chain), which is comparable to the thermal broadening expected at 250 mK. In contrast, no prominent zero-bias peak is apparent in the bulk of the chain (Fig. 1d), where the spectra typically show a set of about 9-10 resonances within the energy window of the host Pb gap. These resonances are asymmetric in spectral weight, but appear to be close to symmetric in energy about zero, suggesting that they are energetically resolved (see Supplementary Materials). The first peak in the spectra above zero bias is typically about ±150-250 µeV, a value consistent with previous measurements using superconducting tips at higher temperatures[23].

In contrast to measurements at 1.4 K, the ZBP at the end of the chain in the mK temperature experiments can reach more than 1.6 times the normal state conductance above the superconducting Pb gap (see Fig. 1d for example). The ZBP scales approximately proportionally with the conductance set by the tunnel junction impedance, reaching a maximum value of about $0.16e^2/h$ obtained with a 250 kΩ junction impedance (see Supplementary Materials). This value is still smaller than the predicted universal conductance of $2e^2/h$ for a MQP[48], suggesting that we are in the sequential tunneling regime and the temperature of our measurements is still large compared to the tunneling energy scale for coupling between the MQP and the STM tip. Nonetheless, the absence of hybridization with other states, judged by the almost thermally limited width of the ZBP, and our observation that the spectral weight of the zero mode is enhanced (relative to the background) at lower temperatures are significant steps in supporting the MQP interpretation of the ZBP in Fe chains. We also note that the ZBP is isolated in energy and position from other Shiba states, so it is not caused by a trivial Shiba state whose energy disperses as the chain diameter tapers near its end[49].

Having established the presence of a sharp ZBP, we examine its spatial structure by comparing zero-bias conductance maps and topographic measurements on the same chain, as shown in Figs. 2a,b. The conductance map confirms that the ZBP is localized at the chain ends, and also reveals an intriguing spatial pattern. The maximum of the ZBP is not centered on the Fe chain, but instead displays twin peaks situated near its sides. This characteristic 'double eye' feature was observed in many different chains (approximately two thirds of those explored); additional examples are presented in the Supplementary Materials. The double eye conductance pattern underscores the importance of tip positioning to detect the strongest ZBP, and may help explain some of the variability in results reported by other groups[39]. Along the chain, the zero-bias conductance typically decays within one nanometer to a characteristic value around 40-50% of the normal state conductance in the middle of the chain, with small oscillations in amplitude (Fig. 2c). This residual zero-bias conductance likely results from a combination of thermally broadened Shiba states and the exponentially decaying tail of the zero-energy end state (see

Supplementary Materials). A cross-section of zero-bias conductance transverse to the wire axis at its end, presented in Fig. 2d, also shows the rapid decay of the zero bias state from the Fe into the Pb substrate. Despite this rapid decay, we show below that the double eye feature results from substantial MQP spectral weight in the Pb atoms adjacent to the Fe chain.

**Modeling the Spectroscopic Properties of Fe Chains**

To understand the spectroscopic features of the Fe chains and their spatial dependence, we perform model calculations that take into account the chain structure, the hybridization of its electronic states with the Pb substrate, and the influence of the STM tip trajectory on the measurements. Previous studies[23] determined that the most likely structure for the atomic Fe chains is a single atom wide, with three atoms stacked vertically in a zigzag structure that is partially embedded between the rows of the Pb(110) surface (Fig. 2c, inset). Following the numerical approach outlined in ref. 24, we compute the spectroscopic properties of the embedded zigzag Fe chain and compare them with the results of our STM measurements (see Supplementary Materials for model parameters).

Figure 3 shows the calculated LDOS as a function of energy and position along a chain of finite length (21 nm), which can be compared with the data in Fig. 1. The model calculations qualitatively match the experimental results: they show a pronounced ZBP that is localized to the end of the Fe chain, as well as substantial subgap spectral weight throughout the chain whose amplitude is electron-hole asymmetric. The observed experimental variation of the spectra along the chain likely results in part from modulation of the Fe atomic chain structure due to incommensurability with the Pb substrate (not included in the model) as well as finite size effects. Nevertheless, the extended profiles of the subgap states along the chain, seen in both experiments and model calculations, strongly suggest that these subgap states are manifestations of thermally broadened van Hove singularities of Shiba bands induced by the Fe chain into the Pb gap. The model calculations also show that while the thermal broadening does not preclude us from observing the ZBP due to the MQP, it does prevent the observation of a hard p-wave gap. This is similar to recently published data on semiconductor nanowires[31], where the hard gap obtained at zero field becomes softer than the one presented here in the presence of a magnetic field—the regime in which the ZBP is observed. Finally, our model reproduces the experimentally observed rapid decay of the ZBP associated with the MQP at the end of the chain.

To understand the origin of the 'double eye' spatial structure of the ZBP, we use a simple three-site model that takes into account the MQP spectral weight on the Fe and Pb sites as well as the influence of the trajectory of the STM tip on the conductance (Fig. 4a). This model, whose details are included in the Supplementary Materials, shows that the conductance at zero energy will have a double peak structure transverse to the chain when the ratio of the LDOS at zero energy for the Fe site [$\rho_{Fe}(0)$] relative to the Pb sites [$\rho_{Pb}(0)$] is less than the ratio of the integral of the LDOS at these sites ($\rho_{Fe}$, $\rho_{Pb}$) over the energy window set by the voltage bias. Such a condition can be satisfied because of several realistic factors. First, because the number of orbitals on the Fe atoms in the normal state is larger than that on the Pb atoms, we generically expect $\rho_{Fe}/\rho_{Pb} > 1$. Second and more importantly, the strong suppression of the local order parameter on the Pb atoms near the Fe chain[50], as well as the adapted structure caused by strong bonding between the Fe chain and its nearby Pb atoms[23], results in a significant enhancement of the subgap LDOS [including $\rho_{Pb}(0)$] in the superconducting substrate adjacent to the chain. Consequently, we find that $\rho_{Fe}/\rho_{Pb} > \rho_{Fe}(0)/\rho_{Pb}(0)$ (see Supplementary Materials) and we therefore expect tunneling into the MQP to be larger on the sides of the chain as opposed to its center. The results of our simulations including these realistic factors are shown in Fig. 4b, and they demonstrate a clear double peak structure similar to the experimental observations. The model described above therefore shows that the double eye spatial pattern observed in experiment results from substantial MQP weight in the host superconductor.

**Buried Fe Chains and MQP Signature in Pb Overlayers**

We further investigate the idea that the MQP can have significant spectral weight in the host superconductor by examining Fe chains covered with a monolayer of Pb (Figs. 5a,b). Encapsulating Fe chains with Pb may also influence the strength of the proximity-induced superconductivity; similar experiments on semiconducting nanowires have very recently been explored[31, 51]. We perform spectroscopic measurements on the buried Fe chains at 1.4 K (Fig. 5c), which show a ZBP localized close to what we determine to be the end of such chains based on a comparison of spectroscopic and topographic maps. The background in-gap conductance measured on the Pb overlayers is typically much lower than that measured on top of exposed chains, and it is also nearly electron-hole symmetric. This electron-hole symmetry is also seen in our model calculations when we examine the spectral properties on the Pb sites below the Fe chain (Fig. 5d), and it is likely a consequence of the basic symmetry of the states in the host superconductor. The suppression of the low energy states except the zero mode in these Pb overlayers suggests that the pairing strength of the Fe chain may be enhanced in this geometry. This possibility, together with the observation of a robust ZBP (although broad in this case because of measurement at 1.4 K), indicates that the buried chain geometry may have some key advantages to future studies of MQPs, especially with regard to braiding. Manipulating MQPs in buried circular chain geometries might provide a clean signal without interference from other Shiba states[52].

**Superconducting Tip Measurements and MQP Electron-Hole Symmetry**

As a final test of the MQP interpretation of our experiment, we perform measurements with superconducting tips to investigate the electron-hole symmetry of the zero-energy state. Previous studies[23, 39] of the Fe/Pb atomic chain system have used superconducting tips to achieve higher spectroscopic resolution and to explore the MQP particle-hole symmetry at temperatures of about 1 K. The measurements showed asymmetric (in amplitude) peaks at a voltage corresponding to the superconducting gap of the tip ($eV = \pm\Delta_{tip}$) as well as other in-gap states at higher energies, and they were interpreted to suggest that some of the spectral weight of the zero-energy mode may result from trivial states[39].

We have extended these measurements to lower temperatures in our dilution refrigerator STM, as shown in Fig. 6. Here we show a linecut of spectra along the wire, positioned so that we cross one of the eyes of ZBP. Whereas previous work at about 1 K found that spectroscopic measurements with superconducting tips show considerably better energy resolution than normal tips, we do not find that this improvement continues at lower temperatures. Instead, the width of the ZBP and other in-gap states resolved with the superconducting tips are somewhat broader (120 µeV, at best) than measurements of the same electronic features with the normal tips (90 µeV, at best). This difference may result from an imperfect tip containing quasiparticle excitations, perhaps due to the amorphous nature or small size of the superconducting apex.

Regardless, spectroscopy with a superconducting tip has been proposed as a way to demonstrate the particle hole-symmetry of the MQP, which should produce peaks with symmetric amplitude at $eV = \pm\Delta_{tip}$, as opposed to trivial Shiba states near zero energy, which could have asymmetric amplitude[39, 45]. As shown in Fig. 6 for one particular tip and Fe chain, we indeed observe two symmetric peaks at $eV = \pm\Delta_{tip}$, a behavior that is distinct from the electron-hole asymmetry of all other in-gap features, especially that associated with a Shiba band at $eV \sim$ 2 meV. While we have found a few combinations of chains and superconducting tips that show symmetric peaks at $eV = \pm\Delta_{tip}$ at the end of the chain, we sometimes also observe non-symmetric behavior (see Supplementary Materials). Further efforts are required to determine the experimental conditions that lead to the symmetric peaks expected for MQPs, and whether the condition of the tip and sample can influence such measurements.

**Concluding Remarks and Outlook**

Performing high-resolution measurements of our atomic chain platform at lower temperatures allows us to place a stringent upper bound of 80 µeV splitting of the ZBP. In addition, the double eye spatial pattern of the ZBP and its robustness to the deposition of a superconducting overlayer demonstrates the enhanced MQP spectral weight in the host superconductor. The detailed correspondence between these data and theoretical modeling, provides critical experimental evidence for the predicted topological nature of superconductivity in this system and the interpretation of the edge bounded ZBPs as signatures of localized MQPs. Further investigations of this platform would, however, advance more rapidly if the structure and magnetic properties of the chains could be manipulated and optimized. For example, the complex in-gap structure of our chains is a consequence of their zigzag structure, which gives rise to multiple Shiba bands[24]. Future efforts in constructing such chains using a STM to perform atomic manipulation may be used to build simpler chains as well as provide the opportunity to maximize the induced superconducting pairing within them. The magnetism in such chains also plays a critical role in the emergent topological superconductivity, and it needs to be studied to determine the phase diagram of superconducting phases in such chains[24, 41]. Finally, using thin film superconducting substrates would provide an opportunity to use a parallel magnetic field as a tuning parameter of the properties of chains or other assemblies of magnetic adatoms on their surface and perhaps to ultimately manipulate MQPs in this platform[52].

**Methods**

Atomically ordered Pb(110) surfaces were prepared from a bulk single crystal by several cycles of argon ion sputtering followed by annealing at 250 ºC. To produce the Fe chains, electron beam evaporation of Fe was performed with the substrate held at approximately 85 ºC, followed by 7 minutes of annealing at 175 ºC. The sample was then gradually cooled over approximately 30 minutes and inserted into the STM. Ultrahigh vacuum conditions were maintained throughout this process.

To grow Pb overlayers on the Fe chain samples described above, we used a commercial Knudsen cell and a customized cooling stage for controlling the sample temperature during evaporation. The Pb evaporator was thoroughly degassed and was calibrated by use of a quartz crystal monitor. The temperature of the cooling stage was stabilized by the liquid nitrogen flow rate and the heater installed in the cooling stage. For the data shown in Fig. 5, we first grew Fe chains on a clean Pb substrate and then cooled the sample to 80 K. The buried Fe chain was prepared by evaporating 5 Å of Pb with an evaporation rate of 0.3 Å/min on the cold sample, followed by annealing at 158 K for 10 minutes on the same cooling stage.

Except where noted, dilution refrigerator experiments were performed using a setpoint bias of $V_{set}$ = -5 mV, a setpoint current $I_{set}$ = 500 pA, and an ac rms excitation of 20 µV. When combined with the 250 mK temperature, this leads to an expected experimental broadening of approximately 90 µeV. Assuming Gaussian peaks with a 90 µeV FWHM, we expect to resolve two split peaks when they are separated by at least 80 µeV.

The data that support the plots within this paper and other findings of this study are available from the corresponding author upon reasonable request.

**Acknowledgments**

The work at Princeton has been supported by ONR-N00014-14-1-0330, ONR-N00014-11-1-0635, ONR-N00014-13-10661, NSF-MRSEC programs through the Princeton Center for Complex Materials DMR-142054, NSF-DMR-1104612, NSF-DMR-1420541, NSF EAGER Award NOA-AWD1004957, DOE DE-SC0016239, Simons Investigator Award, Packard Foundation and Schmidt Fund for Innovative Research, and by Gordon and Betty Moore Foundation as part of EPiQS initiative (GBMF4530). This project was also made possible using the facilities at Princeton Nanoscale Microscopy Laboratory supported by grants through ARO-MURI program W911NF-12-1-0461, DOE-BES, LPS and ARO-W911NF-1-0606, and Eric and Wendy Schmidt Transformative Technology Fund at Princeton. B.E.F. acknowledges financial support from the Dicke Fellowship. M.T.R. acknowledges support from the NSF Graduate Research Fellowship Program.

**Author Contributions**

B.E.F., M.T.R., and I.K.D. performed the dilution refrigerator STM measurements. S.J., Y.X., and I.K.D. conducted the measurements on Fe chains capped with Pb overlayers. J.L., Z.W. and B.A.B performed the theoretical modeling and simulations. All authors contributed to analyzing the data and writing the manuscript.

**Competing Financial Interests**

The authors declare no competing financial interests.

**Figure Legends**

**Figure 1 | Zero-bias end mode in Fe chains on Pb. a**, Three-dimensional rendering of an Fe chain on the Pb(110) surface. Scan area is 50 nm x 50 nm, and the $z$ height range is 7.8 Å. Setpoint bias $V_{set}$ = -50 mV and current $I_{set}$ = 200 pA. **b**, Differential conductance d$I$/d$V$ of the bare Pb(110) surface (blue) and its two-gap fit (green) with $\Delta_1$ = 1.42 meV and $\Delta_1$ = 1.26 meV, at fixed temperature $T$ = 250 mK using a broadening $\Gamma$ = 8 μeV. **c**, Linecut of d$I$/d$V$ along the side of an Fe chain showing a sharp zero bias peak (ZBP) localized to the end as well as several sub-gap Shiba states. The white dashed line marks the end of the Fe chain. **d**, d$I$/d$V$ at the chain end (red) and its average in the middle of the chain (purple; averaged over positions > 2.5 nm).

**Figure 2 | Double eye pattern and spatial dependence of the ZBP. a,b**, Map of zero-bias conductance (**a**) along a chain and its corresponding topography (**b**). Scale bar: 2 nm. **c,d**, d$I$/d$V$ as a function of position along (**c**) and transverse to the Fe chain (**d**). Data in (**c**) are averaged over the 1.6 nm chain width. Inset: schematic of zigzag Fe chain partially embedded in the substrate. In (**d**), the end spectrum (red) is averaged over the 1 nm length of double-eye feature, and mid-chain spectrum (purple) is averaged over the remainder of the chain.

**Figure 3 | Theoretical model of the zero-bias end mode. a**, Simulation of d$I$/d$V$ as a function of position along the chain, whose end is marked by the white dashed line. **b**, Theoretically predicted d$I$/d$V$ at the end (red) and middle (purple) of the Fe chain. A ZBP localized to the end

of the chain and subgap Shiba states that qualitatively resemble the experimental data are apparent.

**Figure 4 | Theoretical model of the double eye spatial pattern of the ZBP. a**, Schematic illustration of the topmost Pb atom of the Fe chain and underlying substrate topography, as determined from DFT calculations, and the tip trajectory during measurements. **b**, Simulation of the expected zero-bias conductance as a function of position. A double eye pattern of the ZBP is apparent, consistent with the experimental observations. The dashed black line corresponds to the theoretically predicted topographic signal within this model. Scale bar: two Pb lattice constants.

**Figure 5 | ZBP in Pb overlayer above Fe chain. a**, Topography of a buried Fe chain. ). $V_{set}$ = 90 mV and $I_{set}$ = 100 pA. Scale bar: 2 nm. **b**, Corresponding schematic showing the chain covered with a single monolayer of Pb. **c**, d$I$/d$V$ at the end of the chain (red), in the middle (blue) and on the bare Pb substrate (black). $V_{set}$ = 5 mV and $I_{set}$ = 700 pA. **d**, Theoretically calculated LDOS of the Pb atoms neighboring the Fe chain at its end (red) and middle (blue), and on the bare Pb substrate (black).

**Figure 6 | Superconducting tip spectra and electron-hole symmetry of the zero-energy end mode. a**, d$I$/d$V$ along the side of a Fe chain showing an end mode at $e|V| = \Delta_{tip}$. The white dashed line marks the end of the chain. **b**, Individual d$I$/d$V$ spectra on the bare Pb (black) and at the end (red) and middle (purple) of the Fe chain. The end spectrum shows electron-hole symmetric peak amplitude at $e|V| = \Delta_{tip}$. **c**, d$I$/d$V$ spectra at each position in (**a**), plotted as a function of $|E|$, with

each curve offset for clarity. The end mode at $e|V| = \Delta_{tip}$ is electron-hole symmetric in amplitude, whereas states at higher energies are not. A weaker peak at $e|V| = \Delta_{tip}$ is also visible in the middle of the chain; its origin is likely the same as the residual spectral weight at zero bias measured in the middle of the Fe chains using normal tips.

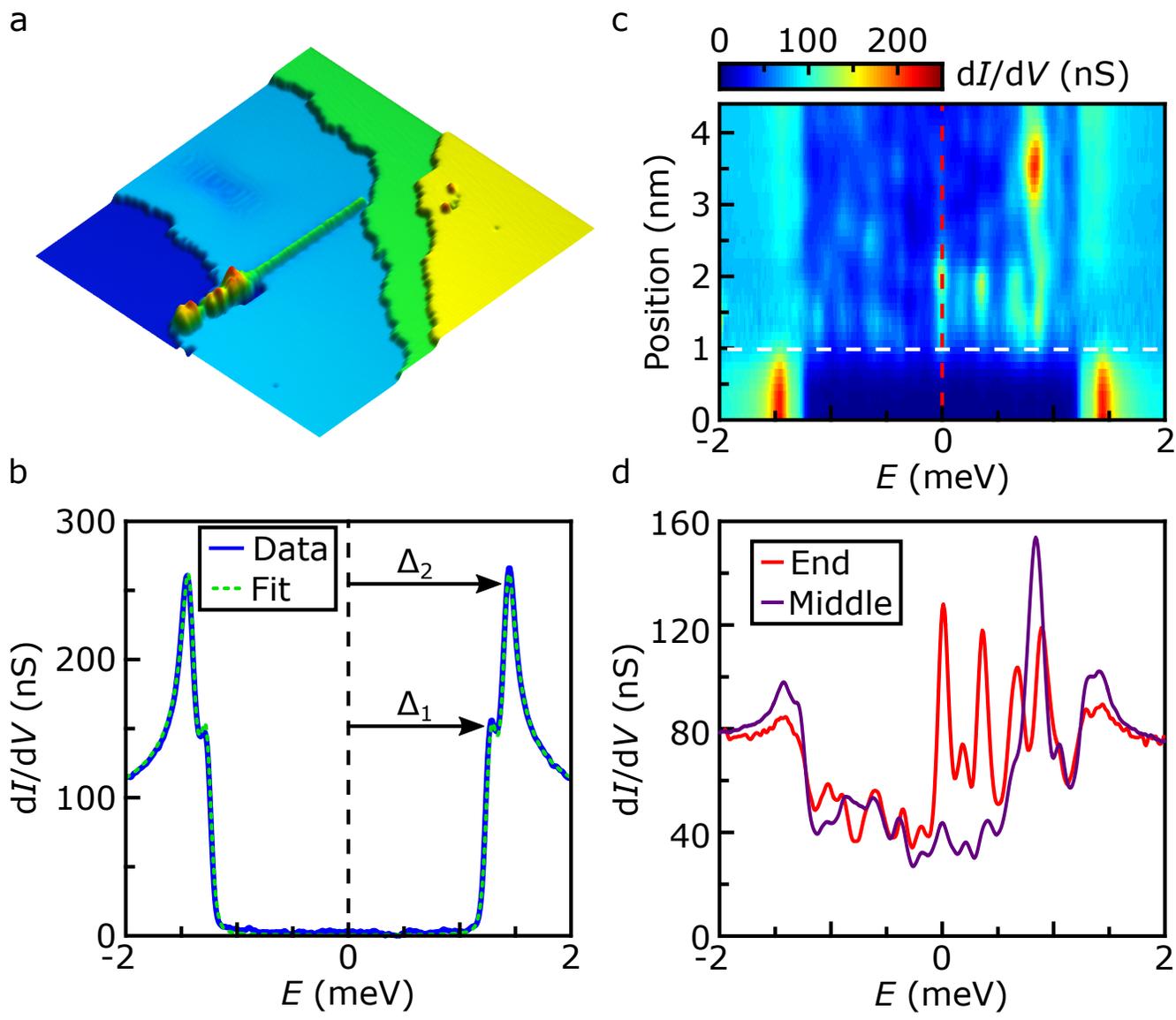

Figure 1

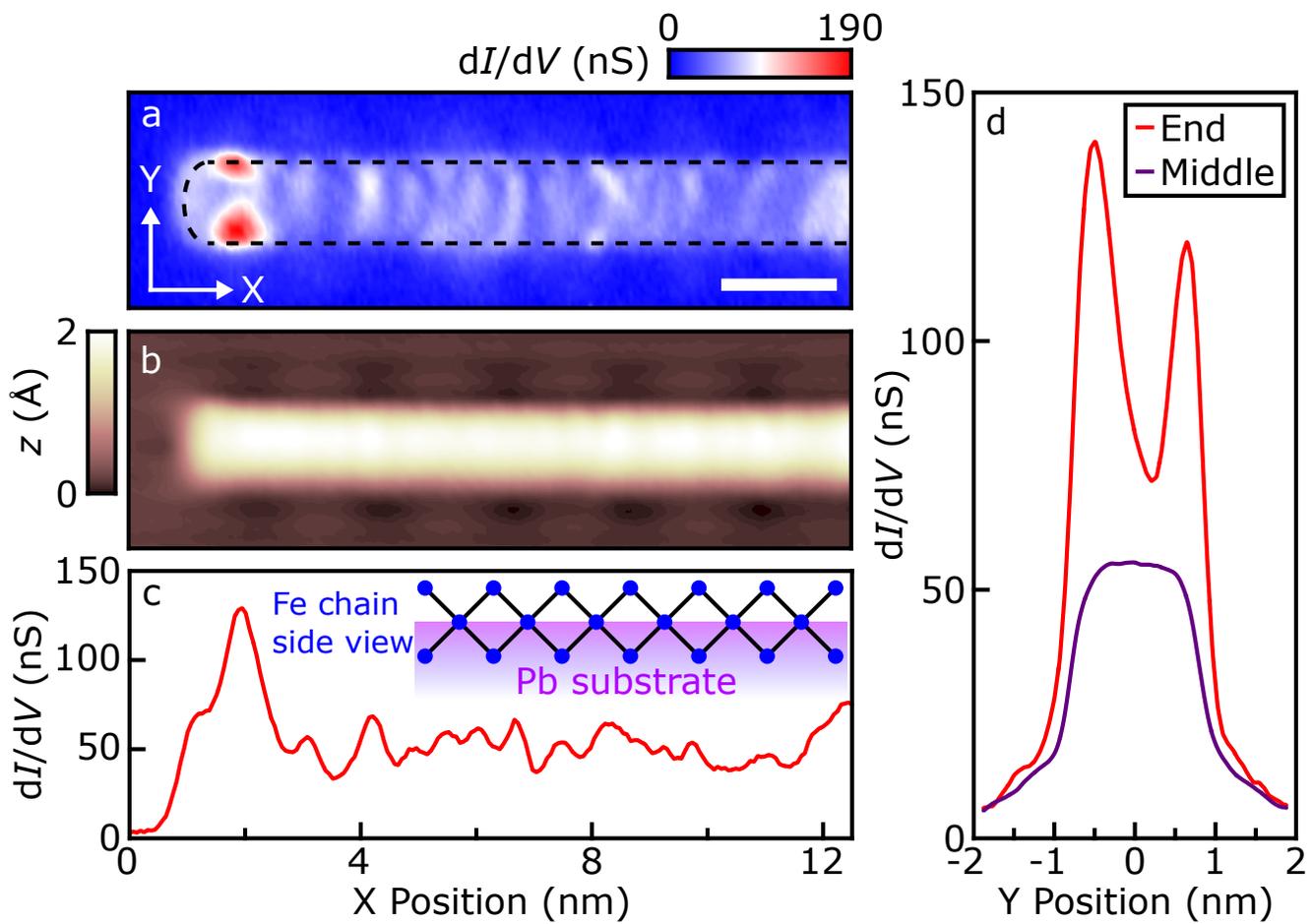

Figure 2

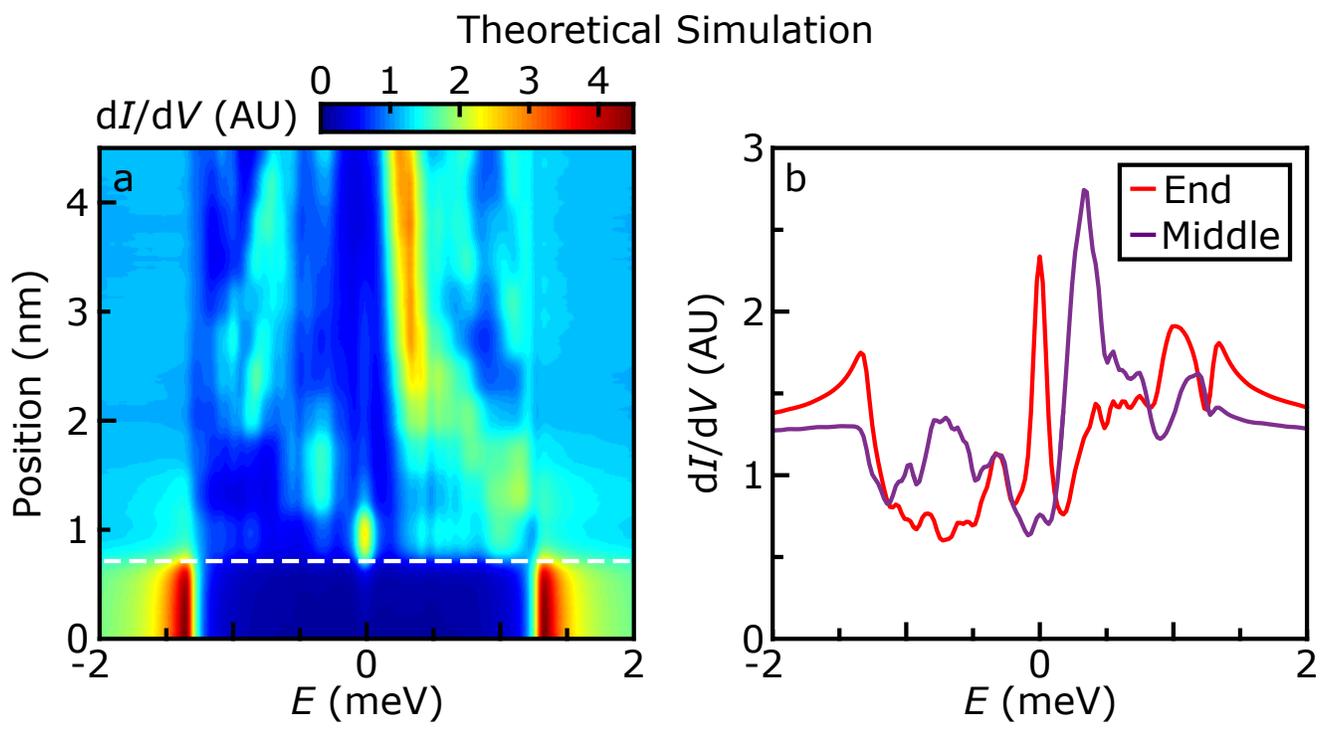

Figure 3

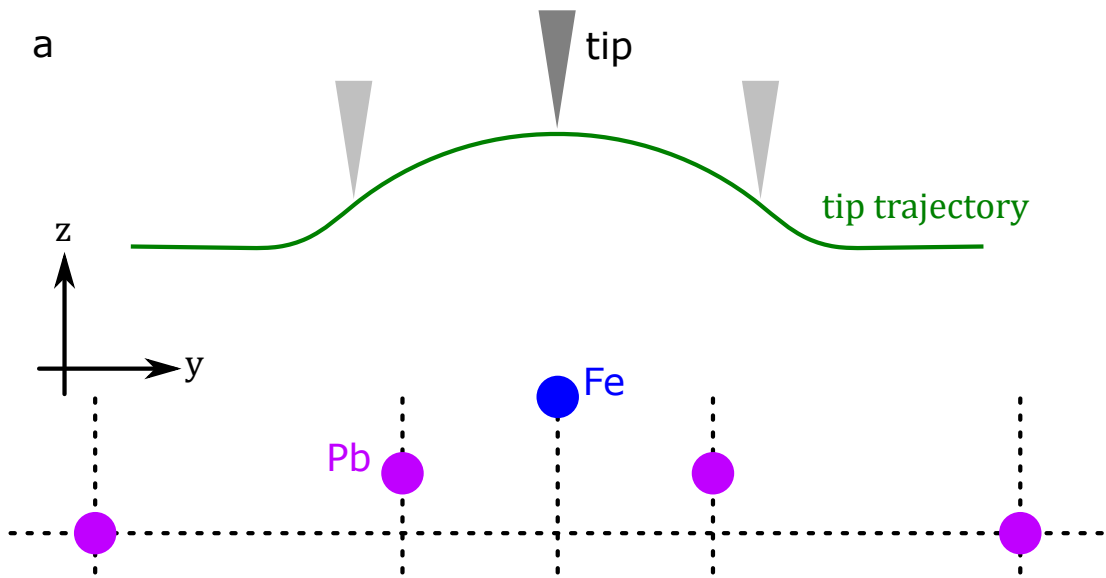

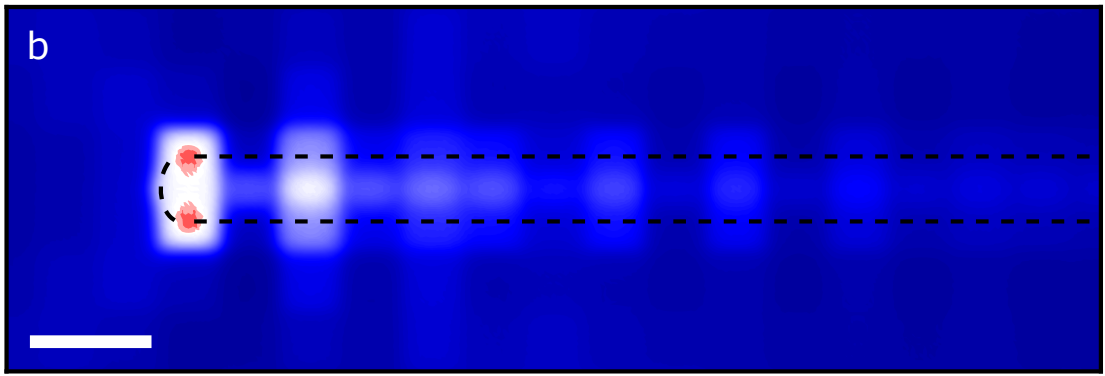

Figure 4

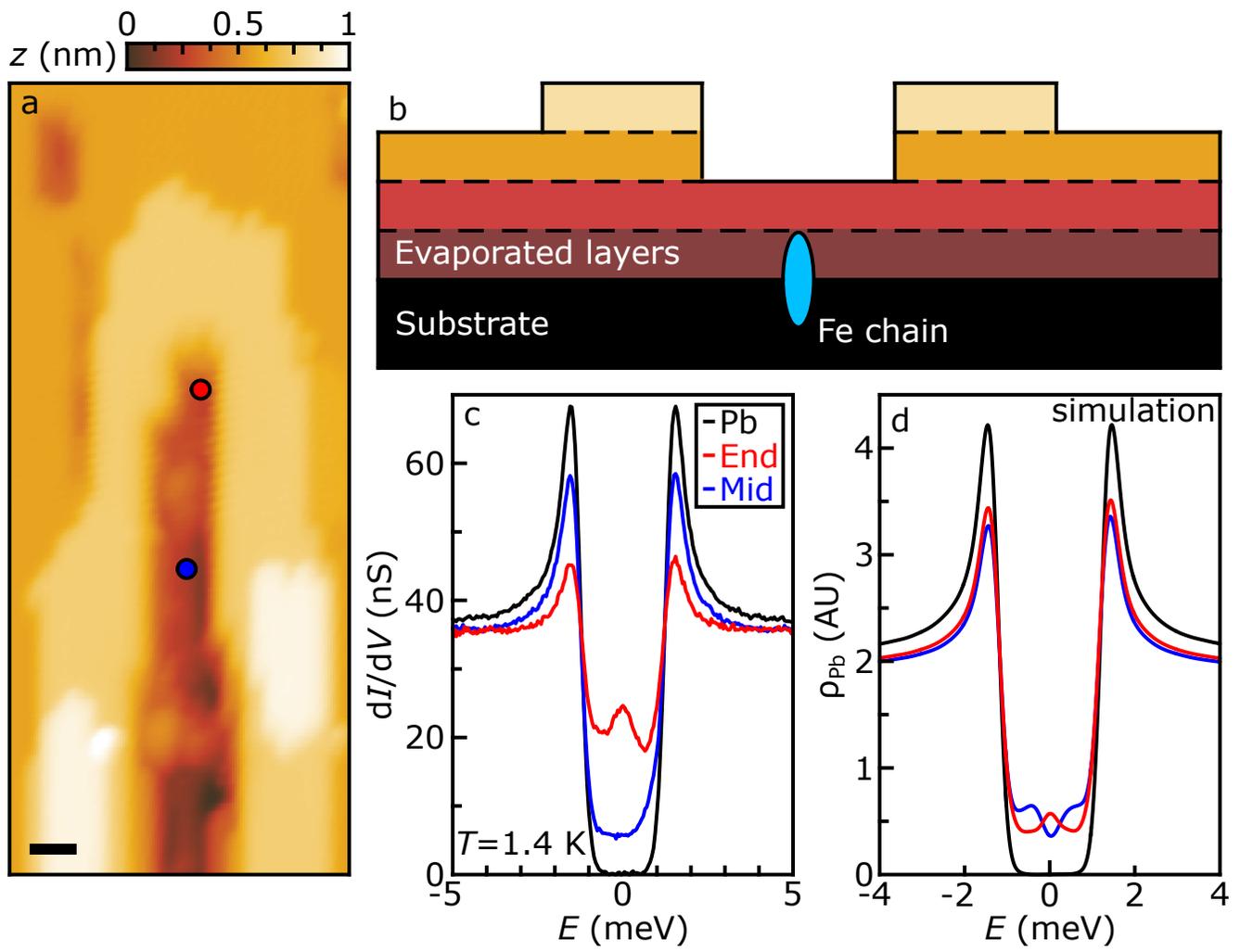

Figure 5

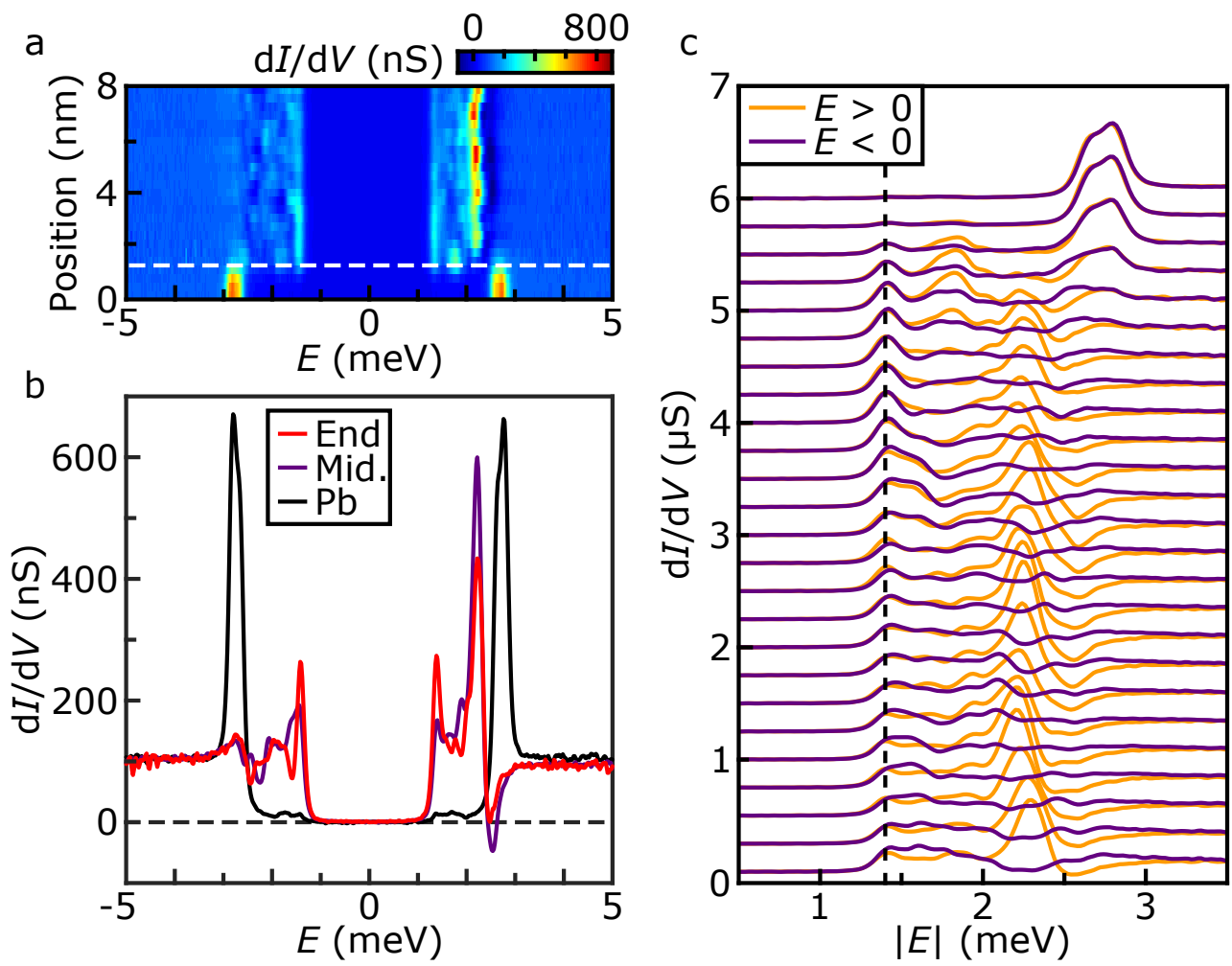

Figure 6

# High-resolution studies of the Majorana atomic chain platform

Benjamin E. Feldman[*], Mallika T. Randeria[*], Jian Li[*], Sangjun Jeon, Yonglong Xie, Zhijun Wang, Ilya K. Drozdov[*†], B. Andrei Bernevig, Ali Yazdani[‡]

*Joseph Henry Laboratories and Department of Physics, Princeton University, Princeton, New Jersey 08544, USA*

[†]Present address: *Condensed Matter Physics and Materials Science Department, Brookhaven National Laboratory, Upton, New York 11973, USA*
[*]These authors contributed equally to the manuscript.
[‡]email: yazdani@princeton.edu

## Supplementary Materials

### Symmetry in energy of the subgap states about zero

To more clearly illustrate that the subgap states at both the end and in the middle of the Fe chains are symmetric in energy about zero, we show in Fig. S1 the spectra from Fig. 1d plotted as a function of $|E|$. The positions of the peaks matches exceptionally well, as expected due to the particle-hole symmetry of Shiba states induced in the superconductor.

Although a weak ZBP is visible in the average mid-chain spectrum shown in Figs. 1d and S1b, this is not a generic feature in all Fe chains. In some locations along the chain whose data is shown in Fig. 1, no ZBP is visible above the background subgap conductance. In addition, we show in Fig. S2 a longer linecut along another Fe chain which also exhibits a ZBP localized to its end, but does not display any ZBP above the background low-energy spectral weight in its mid-chain average (or at local positions in the middle of the chain).

### ZBP conductance as a function of junction impedance

The zero-bias conductance at the end of the chain scales approximately proportionally with the conductance set by the tunnel junction impedance, as illustrated in Fig. S3. As we vary the STM junction impedance $R$ from 10 MΩ to 250 kΩ, the zero-bias conductance increases slightly faster than $1/R$. When the inverse lifetime $\tau$ of the MQP is larger than the tunneling rate $\Gamma$ between the tip and the sample, the differential conductance contributed by sequential tunneling is proportional to $\Gamma/\tau$, whereas that contributed by Andreev reflections is proportional to $(\Gamma/\tau)^2$. The almost linear dependence of the zero-bias conductance on $1/R$, together with the facts that the magnitude of the zero-bias conductance is significantly smaller than $2e^2/h$ and the in-gap conductance is highly asymmetric with respect to zero energy, indicate that the tunneling current in our measurement is dominated by sequential tunneling of single electrons (as opposed to the Andreev tunneling process; see ref. 1). In this regime, we expect that the width of the ZBP is determined by thermal broadening. Changes in the amplitudes of the other sub-gap states are also visible, and likely result from changes in the relative importance of direct and Andreev tunneling mechanisms into Shiba states[1]. The decrease in relative amplitude of the strong Shiba peak at $eV \approx 0.9$ meV is generic and has been observed in multiple chains, but the behavior at other energies is less systematic across different chains. We obtain a maximum zero-bias conductance of about $0.16e^2/h$ with a 250 kΩ junction impedance.

**Additional examples of the zero-bias conductance spatial distribution**

We observe a double eye spatial pattern of zero-bias conductance in a majority of the Fe chains that we have explored, and two additional examples are shown in Fig. S4. The observation of higher zero-bias conductance on the sides of the Fe chain rather than directly above its center is generic, but the detailed shape of the signal does vary in different chains. The asymmetry we sometimes observe in the double eye pattern is likely a result of the detailed STM tip shape and its coupling to the sample. Among the remaining minority of Fe chains that do not exhibit a double eye, we often observe increased zero-bias conductance at their ends, either in the form of a single large peak or multiple smaller ones. A ZBP is also visible in spectra taken at the ends of many of these remaining chains, and an example topography, zero-bias conductance map and end spectrum from such a chain are shown in Figs. S5a-c.

We note that the double eye pattern of the ZBP is not unique to low temperature measurements; similar behavior is also sometimes observed at 1.4 K. However, the improved energy resolution at lower temperature allows us to probe the spatial distribution of the zero-bias conductance more precisely because there is less contribution from thermally broadened subgap states at nearby energies. Another factor that affects our ability to resolve the double eye feature, as well as to detect the ZBP more prominently above the background, is the 'sharpness' of the tip (both in the topography and in terms of the conductance). This is illustrated in Figs. S5d-g, which show two topographies and corresponding zero-bias conductance maps of the same Fe chain taken with two different tips at 1.4 K. The measurement performed with a more blunt tip (Fig. S5e) shows only a single ZBP at the chain end, whereas a sharper tip reveals the characteristic double eye feature (Fig. S5g). The tips that produce the sharpest topography do often resolve the double eye spatial pattern.

**LDOS simulation parameters for Fig. 3**

The simulated differential conductance along the Fe chain in Fig. 3 is performed according to the tight-binding Hamiltonian presented in ref. 2, and it includes thermal broadening of 90 μeV. The model parameters used for the simulation are shown in Table S1. We note that the subgap (Shiba) states that are present in the model calculation are necessarily symmetric in energy about zero due to the particle-hole symmetry of the superconducting host.

However, the spectral density for the electron components, which dictates the measurement in the sequential tunneling regime, is not necessarily symmetric with respect to zero energy.

**Table S1**. Tight binding model parameters for the theoretical simulation presented in Fig. 3.

| Parameter | Value | Parameter | Value |
|---|---|---|---|
| $\mu_{Fe}$ | 1.74 eV | $\epsilon_{Pb}$ | 0.97 eV |
| $\lambda_{Fe}$ | 0.06 eV | $\lambda_{Pb}$ | 0.665 eV |
| $V_{dd\sigma}$ | -0.6702 eV | $V_{pp\sigma}^1$ | 1.134 eV |
| $V_{dd\pi}$ | 0.5760 eV | $V_{pp\pi}^1$ | 0.080 eV |
| $V_{dd\delta}$ | -0.1445 eV | $V_{pp\sigma}^2$ | 0.146 eV |
| $n_{dd\sigma}$ | 3 | $V_{pp\pi}^2$ | 0 |
| $n_{dd\pi}$ | 4 | $V_{pd\sigma}/V_{pd\pi}$ | -2.17 |
| $n_{dd\delta}$ | 4 | $n_{pd\sigma}, n_{pd\pi}$ | 4 |
| $J_{Fe}$ | 2.5 eV | $V_{pd\pi}$ | 0.32 eV |

**Double eye model description**

Owing to the exponential decay of the tunneling current with the distance between the STM tip and the sample, the minimal model to understand the double eye feature involves only three sites on the surface. In this minimal model, the three sites all lie in a plane that is perpendicular to the chain and contains the end of the chain (*i.e.* the $x = 0$ plane). We label the Fe site as 0 and the neighboring Pb sites as ±1 (see Fig. S6 for a detailed schematic). We further denote the positions of the three sites to be

$$(y_0, z_0) = (0, c); \quad (y_{\pm 1}, z_{\pm 1}) = (\pm a, 0). \tag{1}$$

Phenomenologically, the zero-bias differential conductance measured when the tip is parked at $(y, z)$ is given by

$$G(y, z) = \rho_1(0)[w_{-1}(y, z) + w_1(y, z)] + \rho_0(0)w_0(y, z), \tag{2}$$

$$w_i(y, z) = \exp\left[-\sqrt{(y - y_i)^2 + (z - z_i)^2}/\xi_i\right]. \tag{3}$$

Here, $\rho_0(E)$ and $\rho_1(E)$ are the spectral densities at energy $E$ (where the thermal broadening has been taken into account) on the Fe site and the Pb sites (same for sites $\pm 1$), respectively; $w_i$ ($i = 0, \pm 1$) are weight factors that take care of the exponential dependence of the measured current on the distance between the tip and site $i$, with $\xi_i$ the decay length. The position of the tip for the measurement of differential conductance is determined according to a constant total current at a certain bias. Such a total current, similar to Eq. (2), is given by

$$I(y,z) = P_1[w_{-1}(y,z) + w_1(y,z)] + P_0 w_0(y,z), \tag{4}$$

$$P_i = \int_0^{E_c} dE \, \rho_i(E). \tag{5}$$

Here $P_i$ represents an integrated spectral density in an energy range whose cutoff $E_c = eV$ is determined by the applied bias $V$.

In the minimal model, the peaks in the zero-bias differential conductance map (*i.e.* the eyes) correspond to the maxima of $G$ as a function of $y$, under the constraint that fixes the dependence of $z$ on $y$ in Eq. (2) by the constant total current condition

$$I(y,z) = I_0. \tag{6}$$

Explicitly, this constraint determines a trajectory $z_t(y)$ such that

$$\forall y: I(y, z_t(y)) = I_0. \tag{7}$$

We define $b = z_t(0) - c$ and assume $b > a$. Without ambiguity, we will denote $\rho_i = \rho_i(0)$ henceforth.

Using Eqs. (4) and (6), we rewrite Eq. (2) as

$$G(y,z) = \frac{\rho_1}{P_1} I_0 + \rho_1 \left(\frac{\rho_0}{\rho_1} - \frac{P_0}{P_1}\right) w_0(y,z). \tag{8}$$

The above equation is still subject to the constraint of Eq. (6), although the latter has already been used once. Physically, this means that to determine the location of the peaks in the zero bias conductance maps, we must evaluate the maximum of $G$ along the trajectory $z_t(y)$. The rewriting is helpful because it is now sufficient to find the maximum (minimum) of $w_0(y, z_t(y))$ if $\frac{\rho_0}{\rho_1} > \frac{P_0}{P_1}$ ($\frac{\rho_0}{\rho_1} < \frac{P_0}{P_1}$). To this end, we consider a part of the equal-distance-to-$(y_0, z_0)$ contour $z_c(y)$ such that

$$\forall y \in [-a, a]: w_0(y, z_c(y)) = w_0(0, z_t(0)). \tag{9}$$

We now prove $I(y, z_c(y)) \geq I_0$ as follows.

First we note that

$$I(y, z_c(y)) - I_0 = I(y, z_c(y)) - I(0, z_t(0))$$
$$= P_1[w_{-1}(y, z_c(y)) + w_1(y, z_c(y)) - w_{-1}(0, z_t(0)) - w_1(0, z_t(0))]. \tag{10}$$

Then by using the explicit form

$$z_c(y) = \sqrt{b^2 - y^2} + c, \tag{11}$$

we have (assuming, without loss of generality, that $\xi_{-1} = \xi_1 = 1$)

$$w_{-1}(y, z_c(y)) + w_1(y, z_c(y)) \tag{12}$$

$$= \exp\left[-\sqrt{(y+a)^2 + \left(\sqrt{b^2 - y^2} + c\right)^2}\right] + \exp\left[-\sqrt{(y-a)^2 + \left(\sqrt{b^2 - y^2} + c\right)^2}\right] \tag{13}$$

$$\geq 2\exp\left\{-\frac{1}{2}\left[\sqrt{(y+a)^2 + \left(\sqrt{b^2 - y^2} + c\right)^2} + \sqrt{(y-a)^2 + \left(\sqrt{b^2 - y^2} + c\right)^2}\right]\right\} \tag{14}$$

$$\geq 2\exp\left\{-\sqrt{\left[(y+a)^2 + \left(\sqrt{b^2 - y^2} + c\right)^2 + (y-a)^2 + \left(\sqrt{b^2 - y^2} + c\right)^2\right]/2}\right\} \tag{15}$$

$$= 2\exp\left(-\sqrt{a^2 + b^2 + c^2 + 2c\sqrt{b^2 - y^2}}\right) \tag{16}$$

$$\geq 2\exp\left[-\sqrt{a^2 + (b+c)^2}\right] \tag{17}$$

$$= w_{-1}(0, z_t(0)) + w_1(0, z_t(0)), \tag{18}$$

Where the equal sign is valid iff $y = 0$. Therefore $I(y, z_c(y)) \geq I_0$, or, by definition, $z_c(y) \leq z_t(y)$ with the equal sign valid iff $y = 0$. It follows that

$$w_0(y, z_t(y)) \leq w_0(y, z_c(y)) = w_0(0, z_t(0)), \tag{19}$$

which implies that the maximum of $w_0(y, z_t(y))$ occurs precisely at $y = 0$. This further implies that as long as $\frac{\rho_0}{\rho_1} > \frac{P_0}{P_1}$, the maximum of $G(y, z_t(y))$ occurs at $y = 0$. In other words, the double peaks are possible only when $\frac{\rho_0}{P_0} < \frac{\rho_1}{P_1}$. Note that this condition does not involve the exponential decay length $\xi_i$, although the actual position of the double-peaks, when this condition is met, may indeed depend on $\xi_i$.

As we have shown above, the occurrence of the double eye requires the spectral density on the Fe site at the end of the chain ($\rho_0$), rescaled by the background spectral weight on the same site ($P_0$), to be smaller than the spectral density on the nearest-neighboring Pb site rescaled in the same way ($\rho_1/P_1$). On the other hand, the characteristic exponential decay of Majorana zero modes from the end towards the gapped bulk normally leads to a large ratio $\rho_0/\rho_1$ at zero energy in a straightforward simulation ($\rho_0/\rho_1 \sim 10$). Therefore, the solution to the double-eye puzzle must lie in a large ratio $P_0/P_1$, or an enhanced $\rho_1$ at zero energy under realistic circumstances, or a combination of these two factors. A large ratio $P_0/P_1$ is possible considering the number of relevant orbitals in Fe (3d+4s) at the Fermi energy versus that in Pb (6p). This factor alone, however, does not guarantee the double-eye criterion to be satisfied in simulations. Also important are two realistic considerations that may enhance $\rho_1$ significantly. First, the local order parameter in Pb near Fe is strongly suppressed when self-consistency is taken into account[3]. Second, our DFT calculation of the relaxed Fe-Pb hybrid structure indicates that the nearest Pb atoms to the Fe chain are significantly lifted upwards from their regular height on the Pb(110) surface (see Fig. 2F in ref. 4 and the experimentally observed modulation next to the chains in Fig. 2b of this manuscript), such that the coupling between these Pb atoms and the rest of the Pb substrate becomes weaker than its usual strength. These two considerations, although impractical to be included into simulations based on the Slater-Koster tight-binding model that contains all the orbitals, sublattices and dimensions, can be included into a simplified (minimal) model as follows.

The simplified tight-binding model consists of a linear magnetic chain and a two-dimensional superconductor (Fig. S7a), each site including only a single orbital. The pristine Hamiltonian of the simplified model thus reads

$$H = H_{MA} + H_{SC} + H_{MA-SC}, \tag{20}$$

$$H_{MA} = \sum_m \boldsymbol{d}_m^\dagger (M\sigma_z + 2t_{MA} - \mu_{MA})\boldsymbol{d}_m - (t_{MA}\boldsymbol{d}_{m+1}^\dagger \boldsymbol{d}_m + h.c.), \tag{21}$$

$$H_{SC} = \sum_r (4t_{SC} - \mu_{SC})\boldsymbol{c}_r^\dagger \boldsymbol{c}_r \\ + \left[\Delta c_{r\uparrow}^\dagger c_{r\downarrow}^\dagger - \boldsymbol{c}_{r+\hat{x}}^\dagger(t_{SC} - i\,t_{SO}\sigma_y)\boldsymbol{c}_r - \boldsymbol{c}_{r+\hat{y}}^\dagger(t_{SC} + i\,t_{SO}\sigma_x)\boldsymbol{c}_r + h.c.\right], \tag{22}$$

$$H_{MA-SC} = \sum_{<m,r>} t_{MA-SC}\boldsymbol{d}_m^\dagger \boldsymbol{c}_r + h.c., \tag{23}$$

where $\boldsymbol{d}_m^\dagger = (d_{m\uparrow}^\dagger, d_{m\downarrow}^\dagger)$ is the electron creation operator on a magnetic atom site, $\boldsymbol{c}_r^\dagger = (c_{r\uparrow}^\dagger, c_{r\downarrow}^\dagger)$ is the electron creation operator on a superconductor site, $\sigma_{x,y,z}$ are Pauli matrices, and $M, t_{MA}, \mu_{MA}, t_{SC}, \mu_{SC}, \Delta, t_{SO}, t_{MA-SC}$ are the model parameters that have transparent physical meanings. For simplicity, we include the self-consistent local order parameters by setting $\Delta$ on the superconductor sites (blue in Fig. S7a) adjacent to the magnetic chain to be zero, and we reduce the hopping strength between the same set of superconductor sites and the remaining part of the superconductor by a factor of $\gamma$. The spectral function of the hybrid system, with the superconductor being infinite in two dimensions, is then calculated numerically by using standard Green's function techniques (see, e.g. ref. 2). The spectral densities thus obtained are exact in the current model, and have both energetic and spatial resolutions.

Our simulations taking into account both the local order parameter suppression and the reduced hopping strength generally exhibit a strong enhancement of the spectral densities on the superconductor sites adjacent to the magnetic chain, especially on those adjacent to the end of the chain. As one example, for the parameters $M = 1, t_{MA} = 0.2, \mu_{MA} = -0.97, t_{SC} = 1, \mu_{SC} = 3, \Delta = 0.01, t_{SO} = 0.2, t_{MA-SC} = 0.5, \gamma = 0.5$ and an inverse-life-time broadening of value 0.001, the local density of states at the end of the magnetic chain and on its nearest superconductor site are shown Fig. S7b. Moreover, shown in Fig. S7c-f are four maps of the local density of states in terms of both the energy and the position in the direction parallel to the chain. Based on these data, we can further calculate the zero-bias differential conductance map by using essentially the same procedure as described in the previous section but now including all the sites explicitly. Such a zero-bias differential conductance map calculated with parameters $a = 0.5, b = 0.45, c = 0.2, \xi = 0.12$ (see Fig. S6 and Eq. (3); here, $\xi$ is the same for all sites) is presented in Fig. 4b in the main text.

An important ingredient in the above calculation is the local suppression of superconducting order parameter in the Pb sites adjacent to the Fe chain, whose theoretical justification is described above. Observing this effect experimentally, however, is challenging. This is because the order parameter can be locally suppressed without causing a shift in the position of the coherence peaks[5, 6]. Thus, the energies of the coherence peaks

(which do not dramatically change near the Fe chain in our experiment) are not a direct manifestation of the magnitude of the order parameter. Instead, the only experimental method that we are aware of to directly and locally measure the order parameter is through Josephson measurements performed with the STM. Measurements over individual Fe adatoms indeed show that the Pb order parameter is suppressed over an atomic lengthscale around individual Fe atoms without changing the positions of the coherence peaks in the spectra[7].

**Determination of the number of Pb overlayers above buried chains**

We estimate the amount of Pb evaporated in the buried chain experiment based on a calibrated quartz crystal monitor. We assume approximately uniform coverage on the sample, and based on large-scale topographies and the height variations that they exhibit near buried Fe chains, we can determine how many Pb layers cover the chains. Specifically, in our experiment, the crystal monitor showed that we evaporated about 2.5 monolayers of lead. The measured sample surface was nearly uniform (Fig. S8), with only a few small gaps except next to the Fe chains. We can distinguish between exposed and buried chains based on their spectroscopic signature, so we know that the chain is covered with at least one monolayer of Pb. We can then count the number of steps (one) required to reach the uniform height of the surrounding substrate. We also note that in the immediate vicinity of the capped Fe chains, we often observe an annular raised Pb terrace whose height is one monolayer above the background substrate; this likely reflects the tendency of Pb atoms to segregate form the Fe chain instead of forming a perfectly uniform film.

Uncapped Fe chains extend approximately 2 Å above the Pb substrate, which is about the height of a single monolayer of Pb. Therefore, an Fe chain covered by a monolayer of Pb actually means that two monolayers are present above the original substrate. The additional single-layer step required to reach the height of the surrounding substrate then shows that at least three monolayers must have been evaporated in total. This amount is reasonable given the crystal monitor estimate (and the sporadic gaps), whereas four or more monolayers would only be possible if the crystal monitor was off by at least 60%, an unrealistic amount.

**Zero-bias conductance map of a buried chain**

For completeness, we illustrate in Fig. S9 a typical zero-bias conductance map of a buried Fe chain. These data were acquired on the same sample that is shown in Figs. 5 and S8, but at a stage where the sample had been annealed to only 123 K for 10 minutes following Pb overlayer evaporation. This gives rise to a less uniform surface topography (Fig. S9a), but otherwise qualitatively similar conductance behavior. Mapping the subgap conductance on this surface allows us to detect the presence of regions whose shape matches that of the chains. The zero-bias conductance (Fig. S9b) exhibits peaks that are respectively localized to each end of the Fe chain, as expected for MQPs. A reduced amount of residual zero-bias conductance is also present throughout the middle of the chain. In addition, a few mid-chain regions show zero-bias conductance that is as pronounced as at the chain ends; these are likely caused by clumps of Fe atoms that serve to nucleate the chain growth. Previous measurements over such clumps in unburied chains have also shown strongly enhanced subgap conductance, including at zero bias[4].

**Additional examples of measurements with superconducting tips**

Measurements of the Fe chains with superconducting tips have revealed symmetric peaks at $eV = \pm\Delta_{tip}$ in multiple chains, although we sometimes do observe asymmetric behavior. The conductance as a function of position is shown in Fig. S10a for a second chain that shows the symmetric behavior expected to result from a MQP. The conductance on the bare substrate clearly shows the two-gap structure of the Pb, indicating a high-quality superconducting tip, and symmetric peaks at $eV = \pm\Delta_{tip}$ that are well separated from neighboring states are apparent in measurements performed near the end of the chain (Fig. S10b). A full set of spectra are plotted in Fig. S10c as a function of $|E|$, which allows for direct comparison of the electron- and hole-like subgap amplitude. Minimal electron-hole asymmetry is apparent at $eV = \pm\Delta_{tip}$ whereas pronounced electron-hole asymmetry is observed for states at higher energies.

Measurements of some chains, however, do not show electron-hole symmetry at $eV = \pm\Delta_{tip}$, as is the case for data shown in Fig. S11. Although the superconducting tip quality is also good in this measurement and a peak in conductance is present near the end of the chain at $eV = \pm\Delta_{tip}$, Fig. S11c clearly shows that the conductance is asymmetric at all energies and locations. Therefore, we conclude that even if a MQP is present, other nearby subgap Shiba states are also present in this chain. At present, we do not have a detailed theoretical or experimental understanding of which tip and/or chain characteristics give rise to symmetric or asymmetric behavior. This topic represents an interesting area to explore in the future, but is beyond the scope of this work.

**Supplementary references**

**Supplementary Figure Legends**

**Figure S1 | Symmetry of the subgap states about zero energy. a**,**b** Plot of the end (**a**) and mid-chain (**b**) spectra in Fig. 1d as a function of $|E|$. The peaks at positive and negative energies occur at nearly identical positions.

**Figure S2 | Additional example of conductance spectra along another Fe chain. a,** Linecut of d$I$/d$V$ along the side of an Fe chain showing a ZBP localized to the end as well as several sub-gap Shiba states. The white dashed line marks the end of the Fe chain. No ZBP is visible in the middle of the chain. **b,** d$I$/d$V$ at the chain end (red) and its average in the middle of the chain (purple; averaged over positions > 2 nm).

**Figure S3 | Dependence of spectra on junction impedance. a,** Normalized d$I$/d$V$ (relative to STM junction impedance $R$) for $R$ ranging from 250 k$\Omega$ to 10 M$\Omega$, measured directly over one of the double-eye features of an Fe chain. **b,** The ZBP scales approximately with $1/R$, reaching a maximum of $0.16 e^2/h$ for $R = 250$ k$\Omega$.

**Figure S4 | Additional examples of double eye zero-bias conductance. a,b**, Topography (**a**) and zero-bias conductance map (**b**) for another typical Fe chain. **c,d**, Topography (**c**) and zero-bias conductance map (**d**) for a third Fe chain. In both cases, the zero-bias conductance shows two peaks that are localized to the end of the chain in a double eye pattern. Scale bars: 1 nm.

**Figure S5 | Atypical chain example and tip dependence. a,b**, Topography (**a**) and zero-bias conductance map (**b**) showing a single ZBP at the chain end. Scale bar: 1 nm. **c**, Corresponding d$I$/d$V$ spectrum at the chain end showing a ZBP and averaged mid-chain spectrum. **d-g**, Topographies (**d,f**) and zero-bias conductance maps (**e,g**) of the same Fe chain taken at 1.4 K with two different tips. $V_{set}$ = 10 mV and $I_{set}$ = 750 pA for panels (**d**)-(**g**).

**Figure S6 | Schematic of theoretical model parameters.** Detailed schematic of the model that helps elucidate the spatial pattern of the zero bias conductance.

**Figure S7 | Details of the double eye simulation. a**, Sketch of the simplified tight-binding model that includes the likely microscopic mechanisms responsible for the double eye feature. Magnetic atom sites are represented by red solid circles, and superconductor sites are represented by empty circles with those adjacent to the magnetic atoms highlighted in blue. The local order parameters on the blue sites are fully suppressed to account for the self-consistency; the coupling between the blue and the remaining superconductor sites is reduced to account for the structural modification observed in DFT calculations. **b**, LDOS, normalized by its integrated value in the range [-5Δ, 0], at the end of the magnetic chain and on its immediate neighbor in the superconductor. **c-f**, Maps of normalized LDOS as a function of the energy and the longitudinal

position ($X$) along the chain. Here, the four maps correspond to the magnetic atom (MA) chain and three lines of superconductor sites (SC1/2/3) of different distance to the magnetic chain [see panel (**a**)].

**Figure S8 | Large-scale topograph of the buried chain geometry.** Topograph showing an approximately uniform surface height following Pb evaporation and light annealing. The topographic signal indicates that just under 3 monolayers were evaporated in total, with one overlayer covering the Fe chain. $V_{set}$ = 90 mV and $I_{set}$ = 100 pA.

**Figure S9 | Zero-bias conductance map of a buried Fe chain. a**, Topograph showing a buried Fe chain. The increased number of small steps in the surrounding substrate is a result of a lower annealing temperature of 123 K for 10 minutes after Pb overlayer deposition. Scale bar: 5 nm. **b**, Corresponding zero-bias conductance map. The conductance is peaked at the chain end but exhibits some weight throughout the chain. The regions of especially high zero-bias conductance in the middle of the chain are likely a result of larger clumps of Fe atoms, which act as nucleation centers for initial chain growth. $V_{set}$ = 5 mV, $I_{set}$ = 200 pA, and ac excitation $V_{rms}$ = 40 µV.

**Figure S10 | Additional example of superconducting tip measurements. a**, d$I$/d$V$ along the side of a Fe chain showing an end mode at $e|V| = \Delta_{tip}$. **b**, Individual d$I$/d$V$ spectra on the bare Pb (black) and at the end of the Fe chain (red) showing electron-hole symmetric peak amplitude at $e|V| = \Delta_{tip}$. **c**, d$I$/d$V$ spectra from (**a**), plotted individually as a function of $|E|$, with each curve offset for clarity. The states at $e|V| = \Delta_{tip}$ are electron-hole symmetric, whereas those at higher energies are not.

**Figure S11 | A third example of superconducting tip measurements: asymmetric amplitude at $e|V| = \Delta_{tip}$. a**, d$I$/d$V$ along the side of a Fe chain showing an end mode at $e|V| = \Delta_{tip}$. $V_{set} = -3$ mV and $I_{set} = 1$ nA. **b**, Individual d$I$/d$V$ spectra on the bare Pb (black) and at the end of the Fe chain (red) showing electron-hole asymmetric peak amplitude at $e|V| = \Delta_{tip}$. **c**, d$I$/d$V$ spectra from (**a**), plotted individually as a function of $|E|$, with each curve offset for clarity.

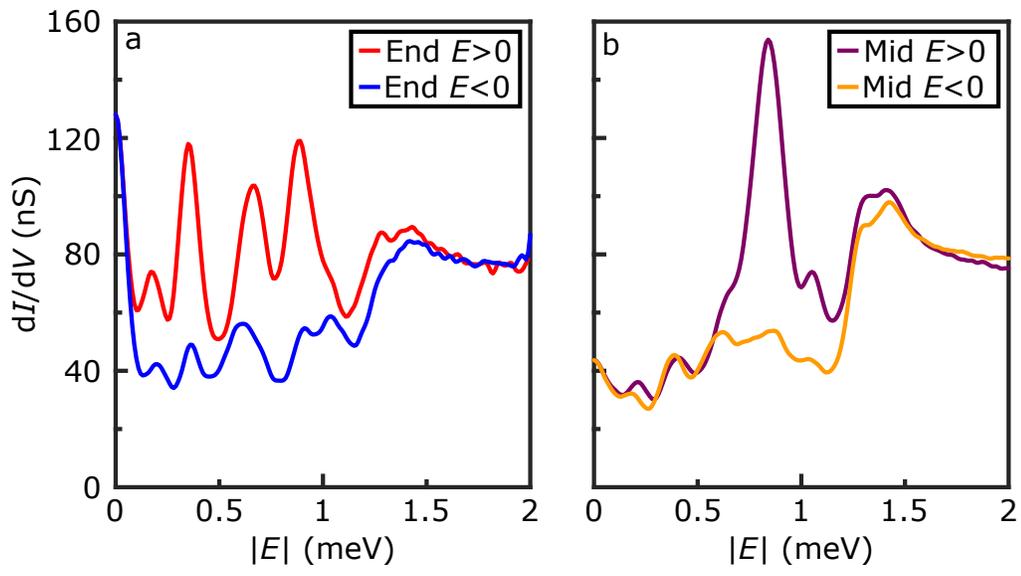

Figure S1

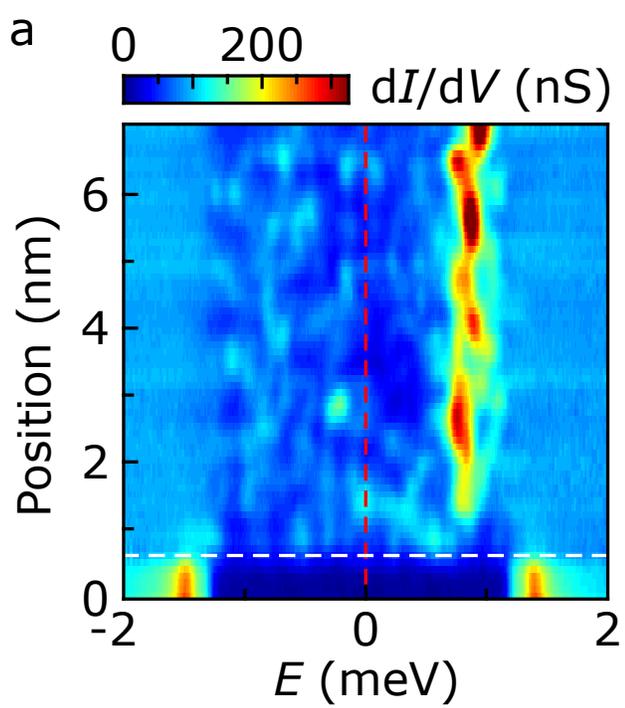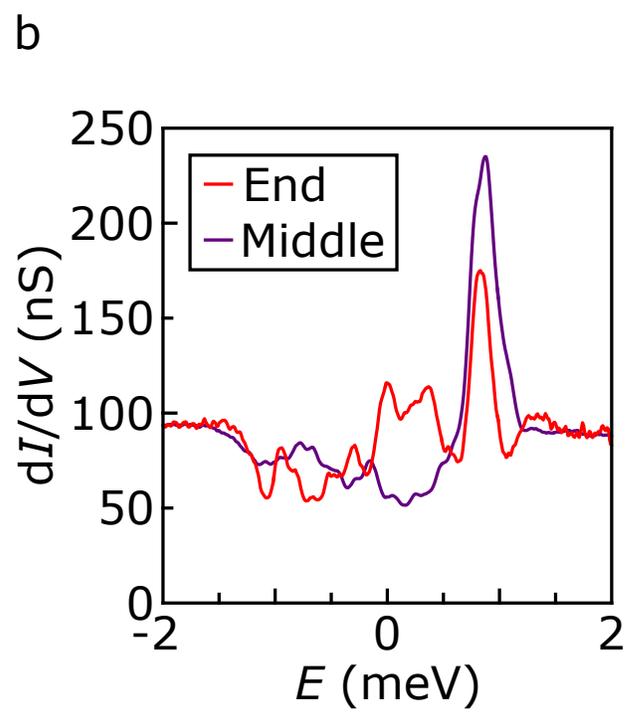

Figure S2

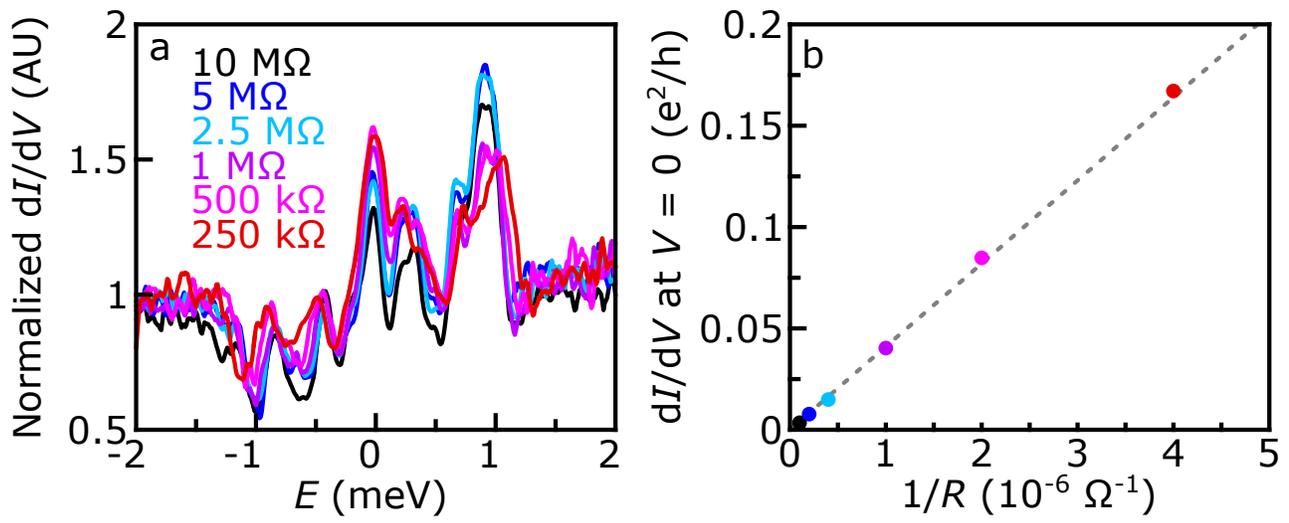

Figure S3

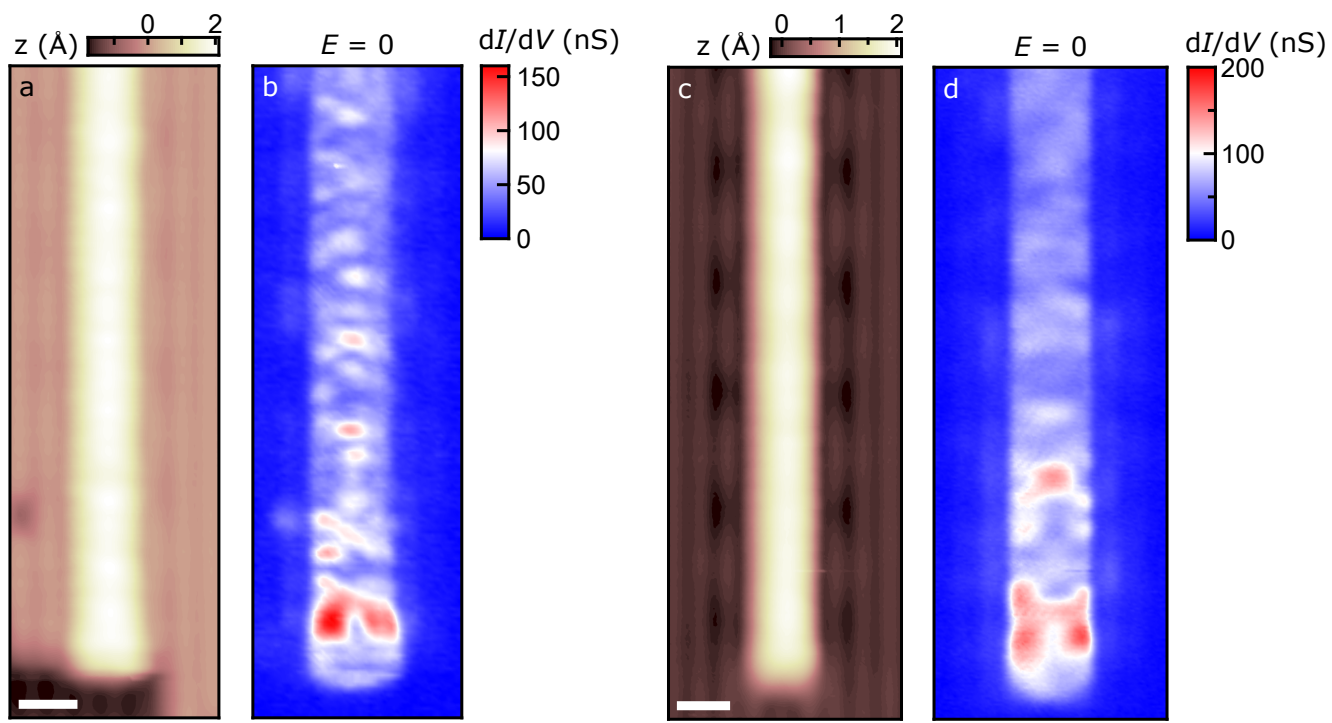

Figure S4

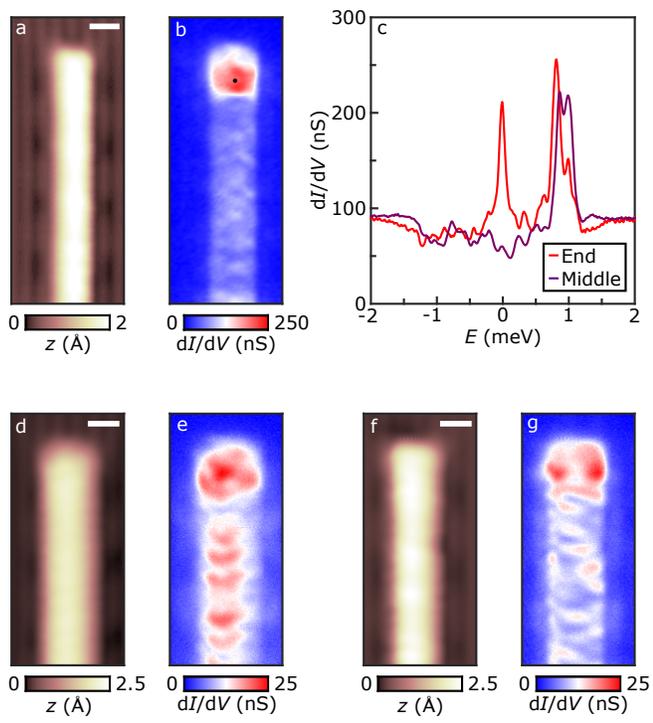

Figure S5

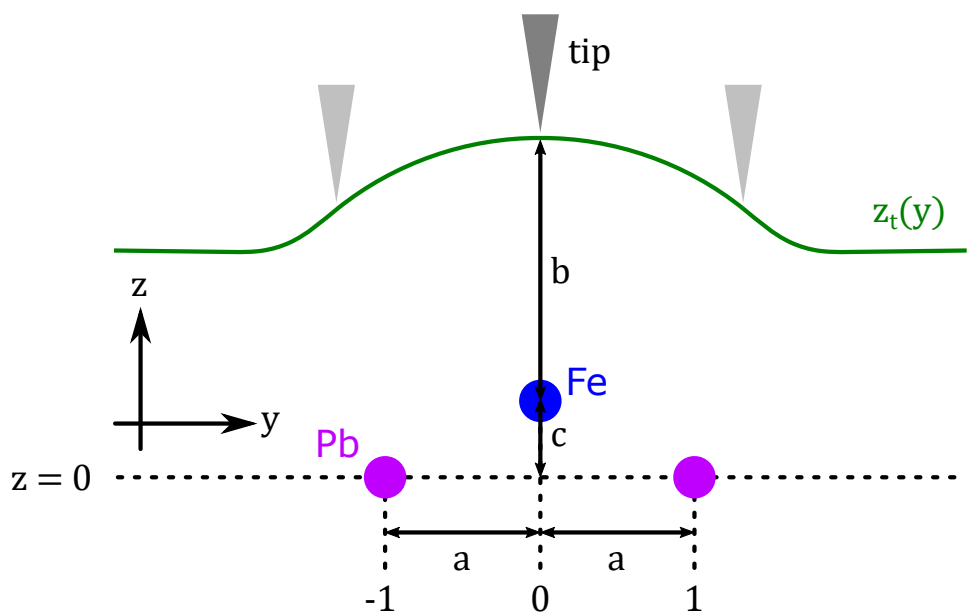

Figure S6

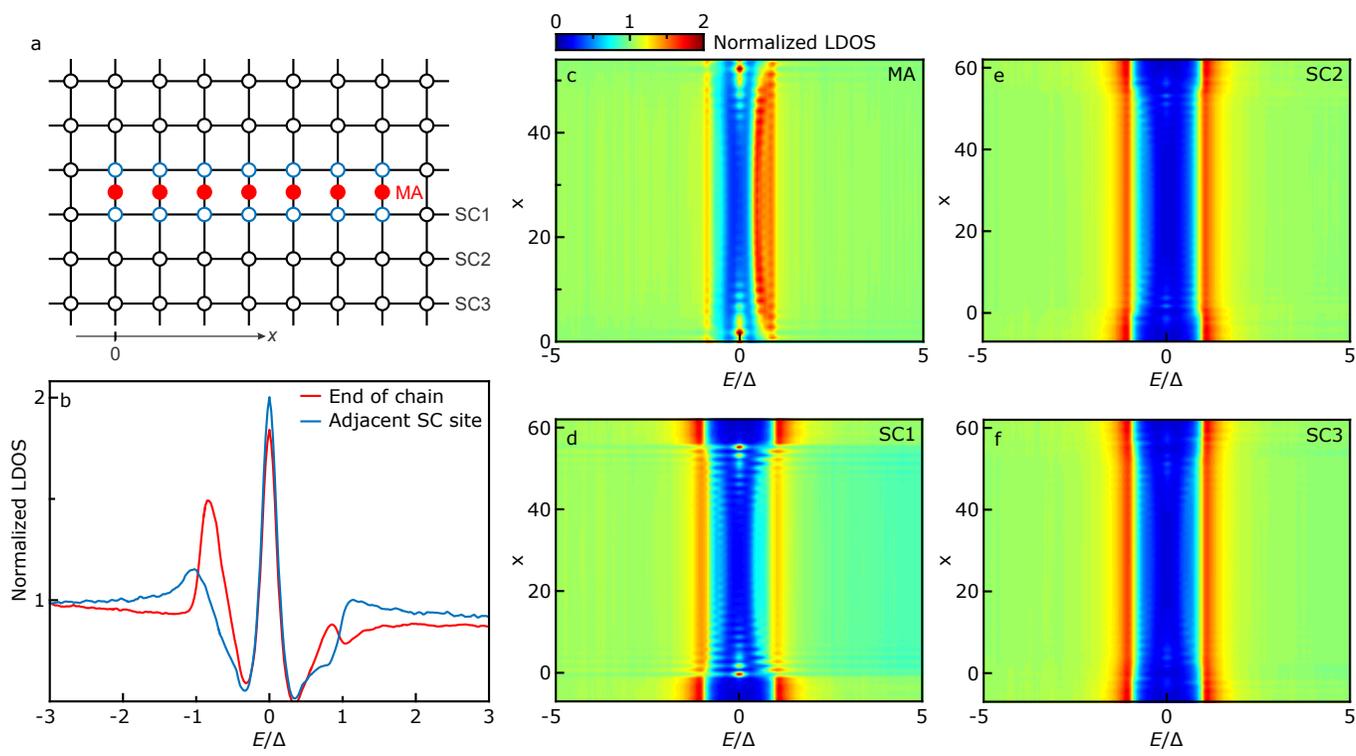

Figure S7

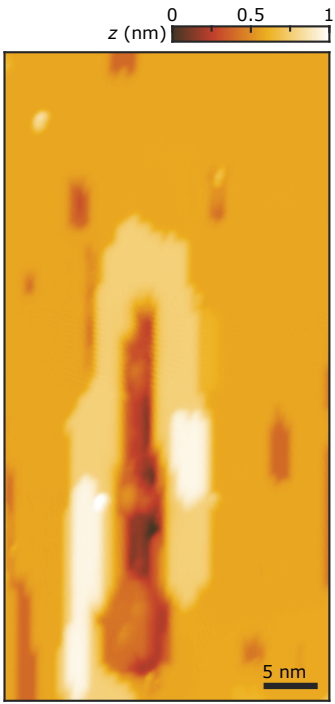

Figure S8

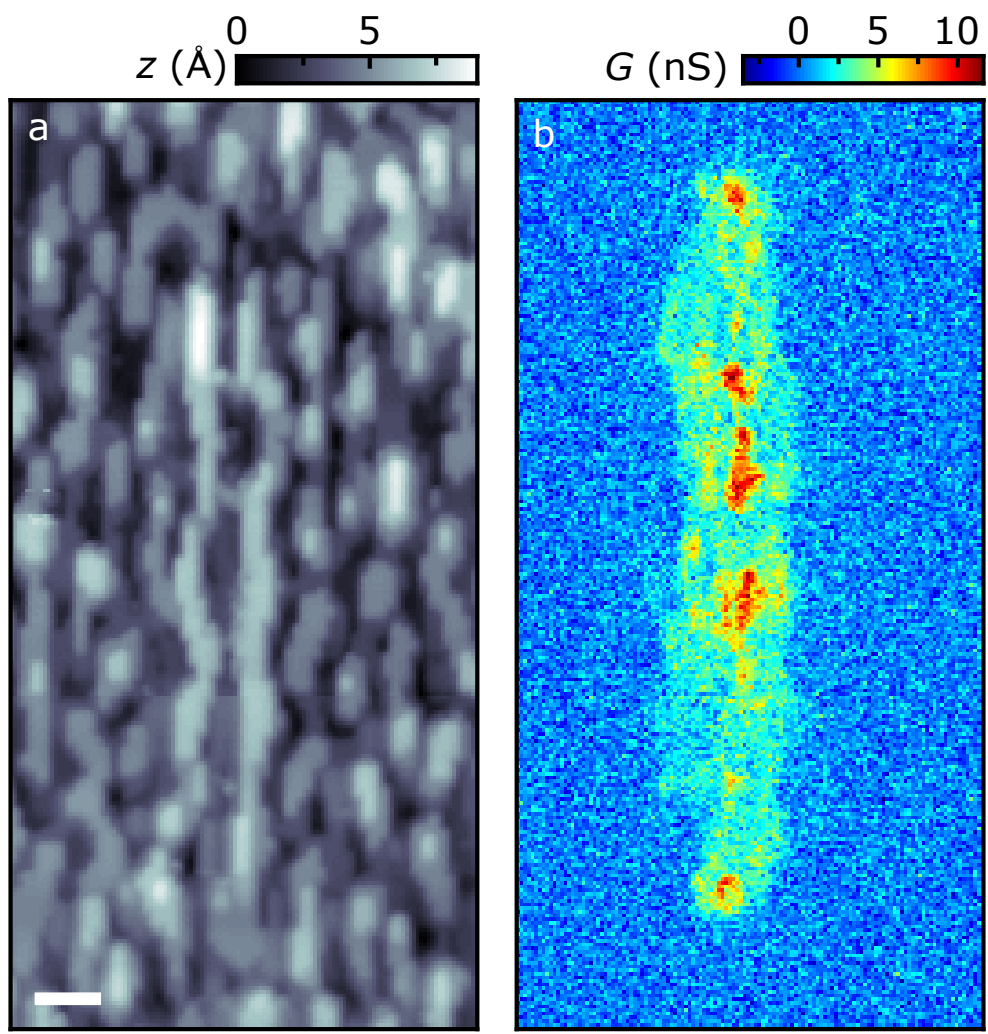

Figure S9

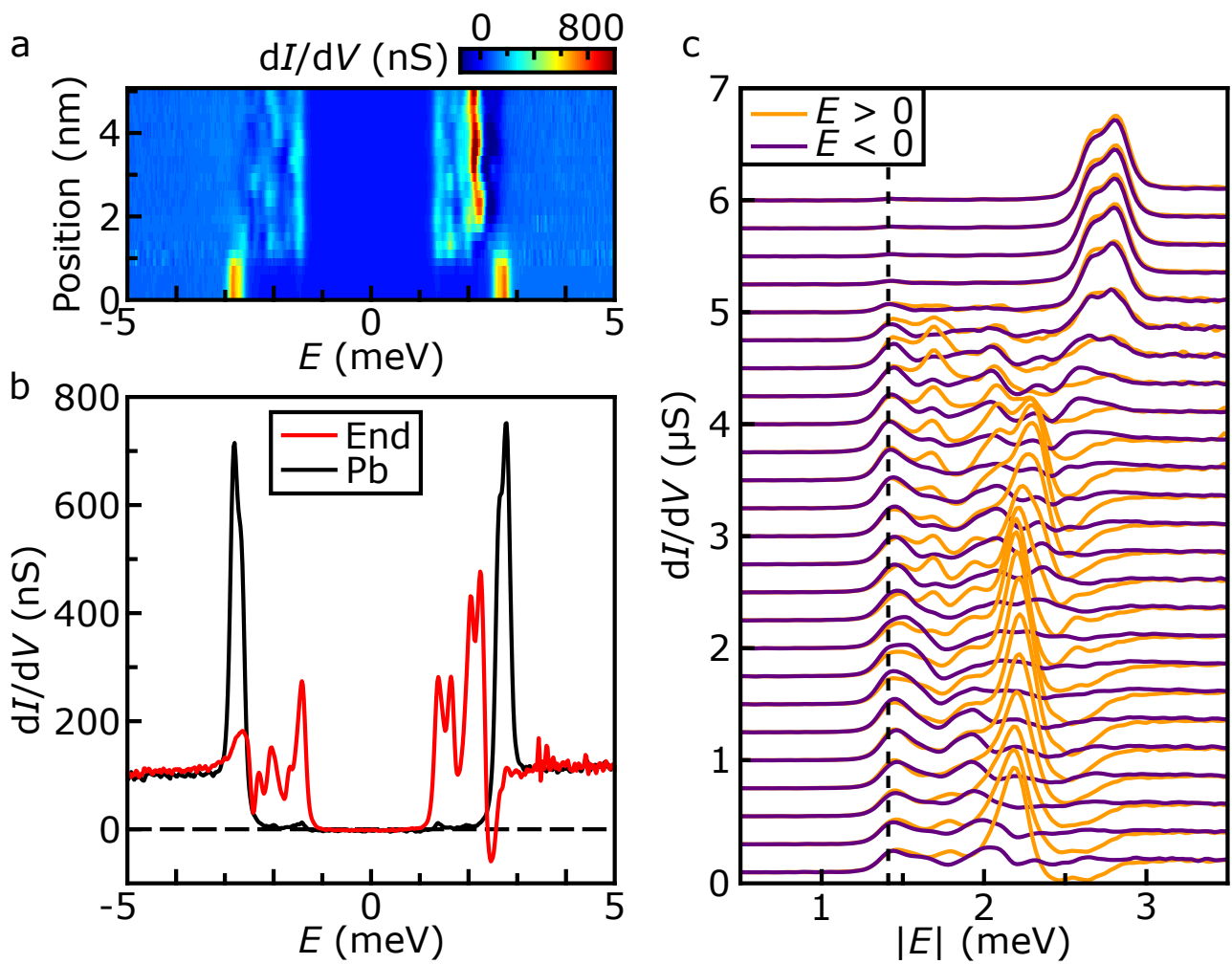

Figure S10

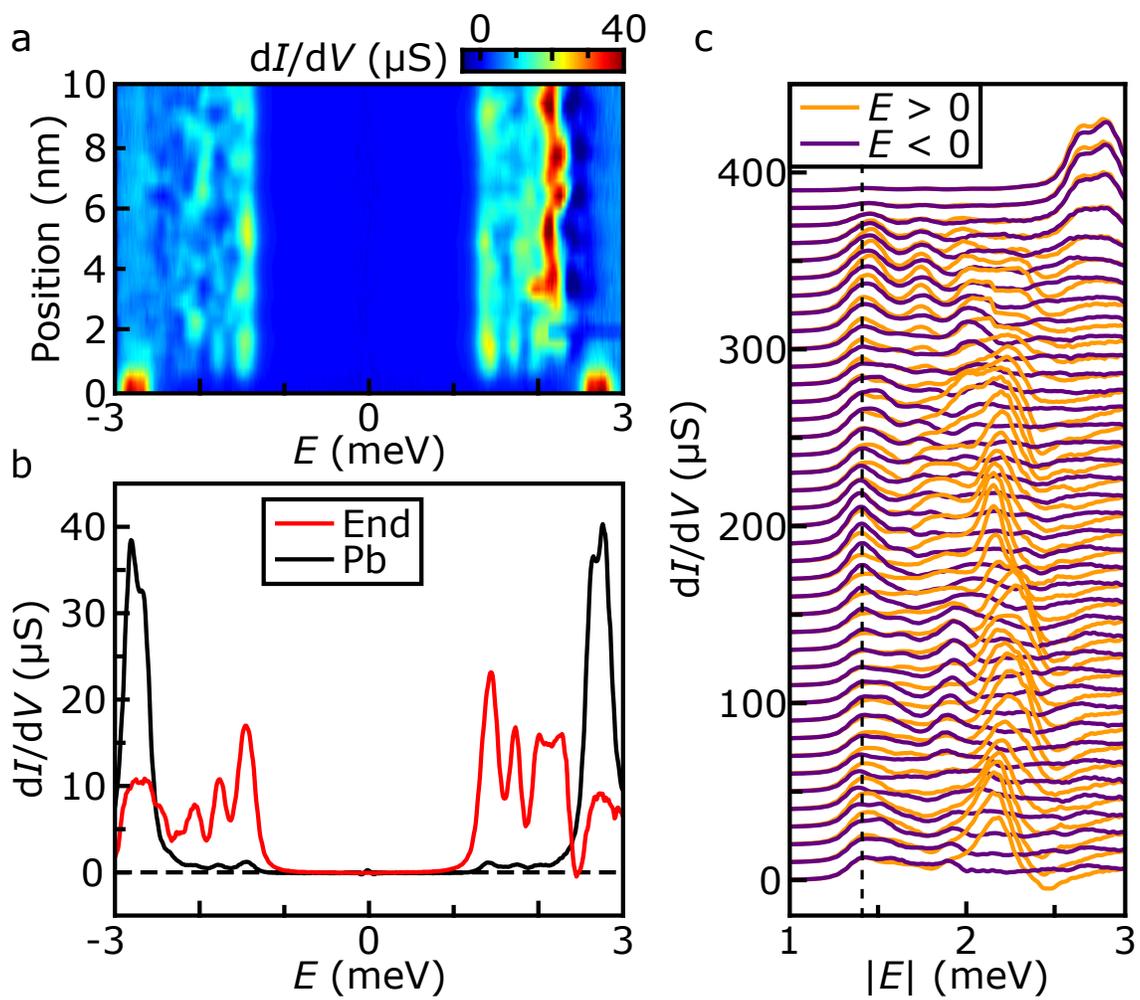

Figure S11